\documentclass[prb,preprint]{revtex4-1} 

\usepackage{amsmath}  
\usepackage{amsfonts} 
\usepackage[dvipdfmx]{graphicx}

\def\dis{\displaystyle} 

\begin{document}

\title{Lambert $W$ function and hanging chain revisited}
\author{Masato Ito}
\email{mito@auecc.aichi-edu.ac.jp}
\affiliation{Department of Physics, Aichi University of Education, Kariya, 448-8542, JAPAN}

\begin{abstract}
In classical physics, calculating the sag of a hanging chain is a problem that has attracted interest. 
This study aims to solve this problem through experiment and theory.
When the length and distance of both the ends of a hanging chain are given,
the length of sag can be expressed by a Lambert $W$ function
or an irrational function within a certain distance. 
Herein, a simple observation of the sag is presented. 
The result obtained is one of the applications of the Lambert $W$ function in
the field of physics.
Though the shape of a hanging chain is well-known to be a catenary,
the calculation of sag possesses richer aspects of mathematical physics.
\end{abstract}

\maketitle 

%
\section{Introduction} 

Experimental demonstration is important in physics. 
However, there are a few experiments based on the mathematical physics.  
This paper provides an example of a simple
demonstration experiment that exhibits the mathematical elements of
classical mechanics.

In classical mechanics, the physics of a hanging chain includes various fields of mathematics, such as infinitesimal calculus, differential equations, 
and variational method.
It is well-known that the shape of a hanging chain (rope or cable) is catenary
(hyperbolic cosine function).
Theoretically, the curve for which the gravitational potential energy acting on the entire chain is minimized is catenary\cite{goldstein}. 
In other words, a catenary can be derived from the equilibrium of forces acting on the line segment of the chain. 
It can be confirmed by observation that the shape of the hanging chain is in accordance with a catenary. 
Thus, the hanging chain is a good pedagogic example that demonstrates the integration of experiment and theory. 
Let us study the method to calculate the length of sag.
Initially, this seems to be a simple calculation but it is not so. 
Given the length and distance of both ends of a hanging chain,
the length of sag is given by solving transcendental equations,
which will be explained later. 

In this paper, it is shown that the sag is expressed by a Lambert $W$ function
or an irrational function with respect to the distance between both ends of the chain under study. 
To compare the theoretical results and experimental data, a demonstration observing the sag position has been proposed.
Interestingly, the Lambert $W$ function appears
in the demonstration experiment of the hanging chain.
In mathematics\cite{corless-1,corless-2}, the Lambert $W$ function appears in combinational theory,
delay time differential equations, and iterated exponentials, among others.
This function is also widely used to solve problems in physics. 

The rest of the paper is organized as follows. 
Section II briefly outlines the definition of the Lambert $W$ function.  
Section III describes the physics of the hanging chain and determines the equations for deriving the sag. 
The horizontal tension of the chain and the asymptotic solutions of the equations are also provided in this section. 
Section IV compares the theoretical results and experimental data and introduces the demonstration experiment for measuring the sag, and the paper is concluded in 
Section V. 
%
%
\section{Brief review of the Lambert $W$ function}

For $x\in \mathbb{R}$, the Lambert $W$ function is defined by
\begin{align}
W(x)e^{W(x)}=x\,.
\end{align}
The plot of $W(x)$ is shown in Fig. 1.
For $-e^{-1}\leq x<0$, $W(x)$ has two branches.
Then the branches satisfying $W(x)\geq -1$ and $W(x)\leq -1$ 
are denoted by $W_{0}(x)$ and $W_{-1}(x)$, respectively.
For $x\in \mathbb{C}$, $W_{k}(x)$ is a multi-branched function 
labelled by an integer $k$.

The Lambert $W$ function has been widely applied in mathematics, computer science, and
engineering\cite{corless-1,corless-2}.
It has also been used in numerical simulation software, such as Maple and Mathematica.
Furthermore, the Lambert $W$ function is often used to solve problems in physics;
it has applications in general relativity\cite{scott}, 
quantum mechanics\cite{scott,valluri},
falling motion of a massive object with air resistance\cite{packel},
calculating the velocity of solar wind\cite{cranmer}, 
the transient phenomenon in an electric circuit\cite{houai}, and
gravity discharge vessels\cite{digilov}, among others.
The transcendental equation is useful for solving various problems in physics. 
The closed-form analytical expression for the solution of the transcendental equation can be obtained by the Lambert $W$ function. If the transcendental equation is given by
\begin{align}
a^{x}=x+b\,,
\end{align}
definition (1) leads to the solution $x$:
\begin{align}
x=-b-\frac{1}{\ln a}W\left(-a^{-b}\ln a\right)\,.
\end{align}
%
\section{Theory of static hanging chain}

Let us reconsider the shape of static hanging
chain in a uniform gravitational field.
Suppose that the infinitesimal line segment of the chain is labelled by $s$, where the
arc length is measured from the origin, as shown in Fig. 2.
Assume that the chain is composed of a material with uniform density. 
Let $\rho$, $T(s)$, and $\theta(s)$ be the chain's mass per unit length, tension, and tangential angle of the infinitesimal line segment $s$, respectively. Thus, Newton's law gives
\begin{align}
&\frac{d}{ds}
\Bigl(T(s)\cos\theta(s)\Bigr)=0\,,\\
&\frac{d}{ds}
\Bigl(T(s)\sin\theta(s)\Bigr)
=\rho g\,,
\end{align}
where $g$ is the gravitational acceleration.
Eq. (4) leads to
\begin{align}
T\cos\theta=T_{\rm h}\,,
\end{align}
where $T_{\rm h}$ denotes the constant horizontal tension.
Substituting Eq. (6) into Eq. (5), the equation determining the shape of the chain is given by
\begin{align}
\frac{d}{ds}\left(\frac{dy}{dx}\right)=\frac{\rho g}{T_{\rm h}}\,,
\end{align}
where $\dis dy/dx=\tan\theta$.
By defining the dimensionless quantity 
\begin{align}
\beta =\frac{\rho gd}{2T_{\rm h}},
\end{align}
the integration of Eq. (7) leads to the shape of the hanging chain,
imposed by the boundary condition of Fig. 2,
\begin{align}
y(x)=\frac{d}{2\beta}\left\{
\cosh\left(\frac{2x-d}{d}\beta\right)
-\cosh\beta
\right\}\,.
\end{align}
The curve is a so-called catenary.
The length $l$ of the chain is computed as
\begin{align}
l=\int^{d}_{0}\sqrt{1+\left(\frac{dy}{dx}\right)^{2}}dx
=\frac{d}{\beta}\sinh\beta\,.
\end{align}
The ratio of $d$ and $l$ is
\begin{align}
\frac{d}{l}&=\frac{\beta}{\sinh \beta}\,,
\end{align}
and Eq. (9) leads to the ratio $h$ and $d$:
\begin{align}
\frac{h}{d}&=\frac{1}{\beta}\sinh^{2}\left(\frac{\beta}{2}\right)\,.
\end{align}
From Eqs. (11) and (12), the ratio of $h$ and $l$ is given by
\begin{align}
\frac{h}{l}=\frac{1}{2}\tanh\left(\frac{\beta}{2}\right)\,.
\end{align}
As shown in Eq. (13), the value of sag $h$ can be calculated if the value of $\beta$ is known. 
However, the aim of this study was to calculate the value of sag $h$
without knowing $\beta$ (the value of $\rho$ and $T_{\rm h}$).
Since the sag $h$ is experimentally measured while varying $d$,
$h$ should be expressed as function of $d$.

The relation between $d/l$ and $h/l$
is obtained by eliminating $\beta$ from Eqs. (11) and (13).
Eq. (11) is a transcendental equation with respect to $\beta$.
Accordingly, Eq. (13) cannot yield the analytical expression as
$h=h(d)$.
However, the ratio $d/l$ can be expressed as a function of $h/l$ by substituting Eq. (13)
into Eq. (11):
\begin{align}
\frac{d}{l}=
\frac{\dis 1-\left(2h/l\right)^{2}}{\dis 2h/l}
{\rm arc}\tanh\left(\frac{2h}{l}\right)
\,.
\end{align}
Eq. (14) is an implicit equation for $h$ and would be useless in the demonstration experiment.
To observe the explicit behavior of $h$ for $d$, the
asymptotic expression of Eq. (14) is derived with respect to
the dimensionless $\beta$, as defined in Eq. (8).

When $\beta\ll 1\;(\rho gd\ll T_{\rm h})$, i.e. the chain is almost straightly
stretched, Eqs. (11) and (13) are expanded in power series of $\beta$
\begin{align}
\frac{d}{l}\sim &\; 1-\frac{\beta^{2}}{6}\,,\\
\frac{h}{l}\sim &\; \frac{\beta}{4}\,.
\end{align}
Eliminating $\beta$, the dimensionless sag $h/l$ is expressed as
an irrational function
with respect to the dimensionless distance $d/l$:
\begin{align}
\frac{h}{l}&\sim \sqrt{\frac{3}{8}
\left(1-\frac{d}{l}\right)}\,.
\end{align}
Furthermore, Eq. (15) yields the ratio of the horizontal tension $T_{\rm h}$
and mass $m$ of the chain as a function of $d/l$
\begin{align}
\frac{T_{\rm h}}{mg}\sim \frac{\dis\frac{d}{l}}{\dis\sqrt{24\left(1-\frac{d}{l}\right)}}\;,
\end{align}
where $m=\rho l$.
Taking the limit of $d\to l$, the horizontal tension goes to infinity; this indicates that the chain can be straightly stretched with arbitrarily large tension, unless the chain breaks. 

When $\beta\gg 1\;(\rho gd\gg T_{\rm h})$, i.e. the chain is loosely stretched,
Eq. (11) leads to
\begin{align}
\frac{d}{\ell}\sim &\; 2\beta e^{-\beta}\,.
\end{align}
From Eq. (1) and (19), $\beta$ is expressed in terms of
the Lambert $W$ function as
\begin{align}
\beta\sim -W_{-1}\left(-\frac{d}{2l}\right)\,,
\end{align}
where $W_{-1}$ is selected because $\beta\to \infty$ as
$T_{\rm h}\to 0$.
Substituting Eq. (20) into Eq. (13), the dimensionless sag can be expressed as
a function of the dimensionless distance
\begin{align}
\frac{h}{l}\sim \frac{1}{2}\tanh\left(-\frac{1}{2}
W_{-1}\left(-\frac{d}{2l}\right)\right)\,,
\end{align}
and the horizontal tension is given by
\begin{align}
\frac{T_{\rm h}}{mg}=\exp\left(
W_{-1}\left(-\frac{d}{2l}\right)
\right)\,.
\end{align}
Consequently, given the length $l$,
the relation between $h$ and $d$ corresponds to  
Eq. (17) and Eq. (21) for $\beta\gg 1$ and $\beta\ll 1$,
respectively.
%
\section{Measurement of the hanging chain and its mathematical aspect}

By measuring the sag $h$ while
varying the distance $d$ of both ends of the hanging chain,
the measurement graph of $d$ and $h$ and the theoretical
graph can be compared.

To perform the hanging chain demonstration in a classroom, a simple
experiment was prepared, as shown in Fig. 3.
One end of a chain was fixed, while the other end was left free to move on the rail; graph paper was placed behind the hanging chain.
Since the position of the movable end was $d$, the sag $h$ of
the chain could be observed by measuring the position of the middle point of chain.
The chain used in the demonstration was an aluminum mantel chain of length $l = 40$ cm.
In Fig. 4, the dots are the measurement data (dimensionless $d/l$ and $h/l$).
The theoretical curve of Eq. (14) was drawn using Mathematica.
Obviously, the theoretical curve agrees with the measurement data.
The asymptotic equations, Eqs. (17) and (21), are more suitable than the implicit equation, Eq. (14), when teaching the mathematical physics aspect of the hanging chain in a classroom. 

The measurement data and graphs of Eqs. (17) and (21)
are shown in Fig. 5.
At this stage, the relative difference between the theoretical value and measurement data is taken to be less than approximately 1\%. 
As a result, Eq. (17) and Eq. (21) are valid for $0.76<d/l\leq 1$ and
$0\leq h/l<0.63$, respectively.
Consequently, the dimensionless sag is asymptotically given by
\begin{align}
\frac{h}{l}=
\begin{cases}
\dis \frac{1}{2}\tanh\left(-\frac{1}{2}
W_{-1}\left(-\frac{d}{2l}\right)\right)
&\dis 0\leq \frac{d}{l}< 0.63\\
& \\
\dis \sqrt{\frac{3}{8}
\left(1-\frac{d}{l}\right)} &\dis 0.76< \frac{d}{l}\leq 1
\end{cases}\,.
\end{align}

Using Eq. (23), a simple demonstration of the mathematical
aspect of the hanging chain can be performed.
In the experiment of Fig. 3, the movable point was initially placed at the
left edge, and then the point was slowly pulled toward the right edge.
Then, the trajectory of the middle point of the chain could be observed.
On the contrary, the trajectory corresponds to the plot of 
$\dis\left(\frac{d}{2},h\right)$ of Eq. (23).
As shown in Fig. 6, by sticking the plots of Eq. (23) (drawn by Mathematica) behind the
chain, it can be confirmed that the trajectory of the middle point is in agreement with 
the theoretical plots.
In a physics class demonstration, students can see that
the actual trajectory is in agreement with the theoretical curve predicted by
Eq. (23).
%
\section{Conclusions}

Herin, it was shown that the problem of calculating the sag of a hanging chain has great mathematical relevance in classical mechanics. 
Given the length $l$ of the chain, the sag $h$ and distance $d$ can 
be determined by Eqs. (11) and (13) for all the values of $\beta$, which depends on 
the horizontal tension of the chain.
The asymptotic solutions for the two equations are found to derive $h$ as a function of $d$. When stretching the chain almost straightly, as shown in Eq. (17),
the sag is expressed by an irrational function with respect to the distance.
On the contrary, as shown in Eq. (21), when stretching the chain loosely,
the sag is expressed in terms of the Lambert $W$ function, which is applied to
the solution of the transcendental equation.
To determine the range, where the asymptotic behaviors are valid,
the sag was measured via the experiment shown in Fig. 3.
The results are given by Eq. (23).

%
%

%
\begin{figure}[ht]
\begin{center}
\includegraphics[width=7cm]{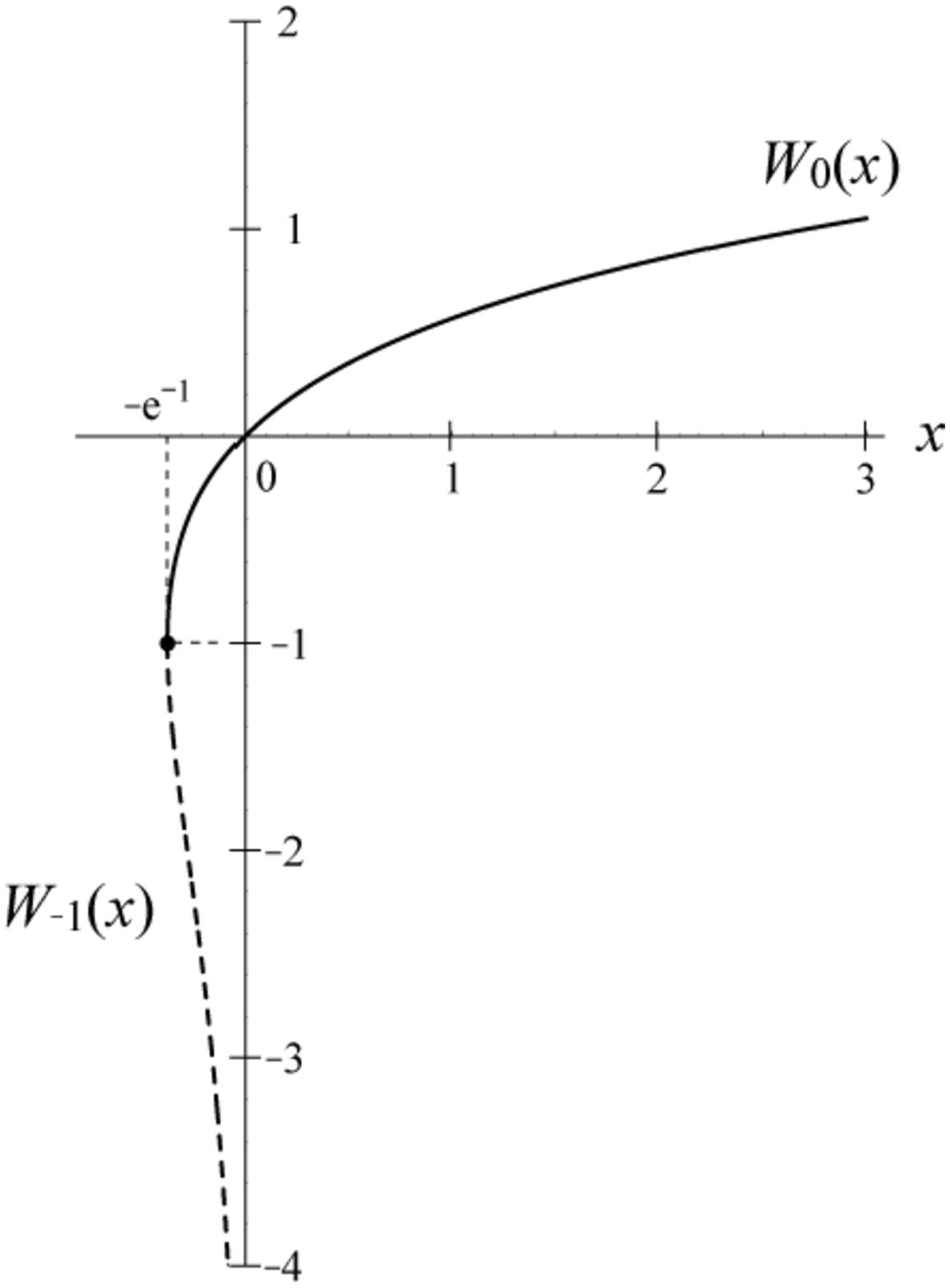}
\end{center}
\caption{Plot of the Lambert $W(x)$ function
for $x\in\mathbb{R}$. The two branches, $W_{0}$ and $W_{-1}$, are shown.}
\end{figure}
\begin{figure}[ht]
\centering
\includegraphics[width=7cm]{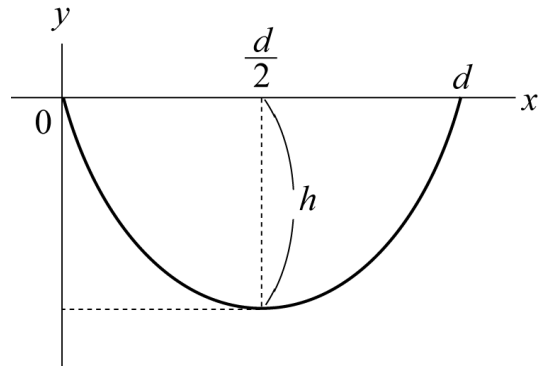}
\caption{Configuration of the static hanging chain.
The positions of both ends are at the same height.
The sag $h$ of the chain is depicted as above.}
\end{figure}
\begin{figure}[ht]
\begin{center}
\includegraphics[width=13cm]{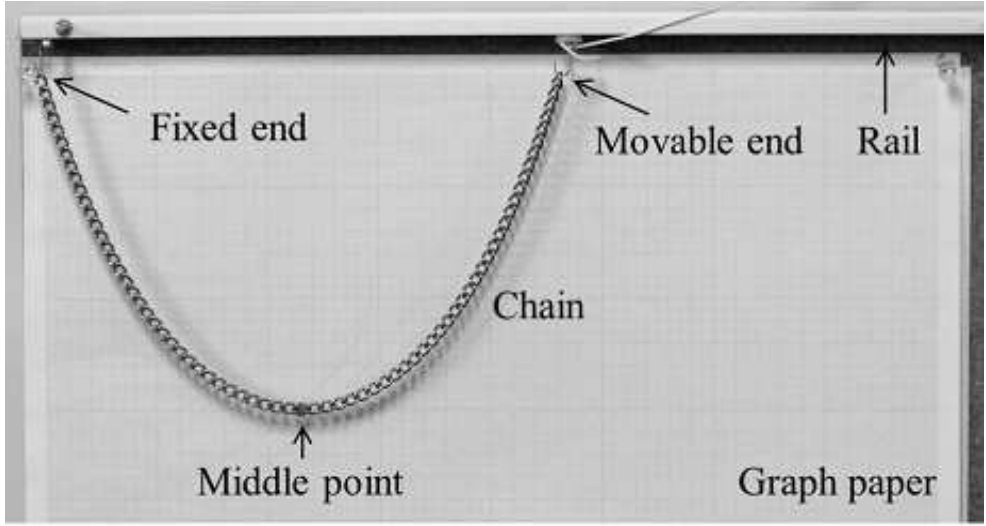}
\end{center}
\caption{Experiment to measure the sag $h$ while varying the distance
between both ends of the chain.}
\end{figure}
\begin{figure}[ht]
\begin{center}
\includegraphics[width=12cm]{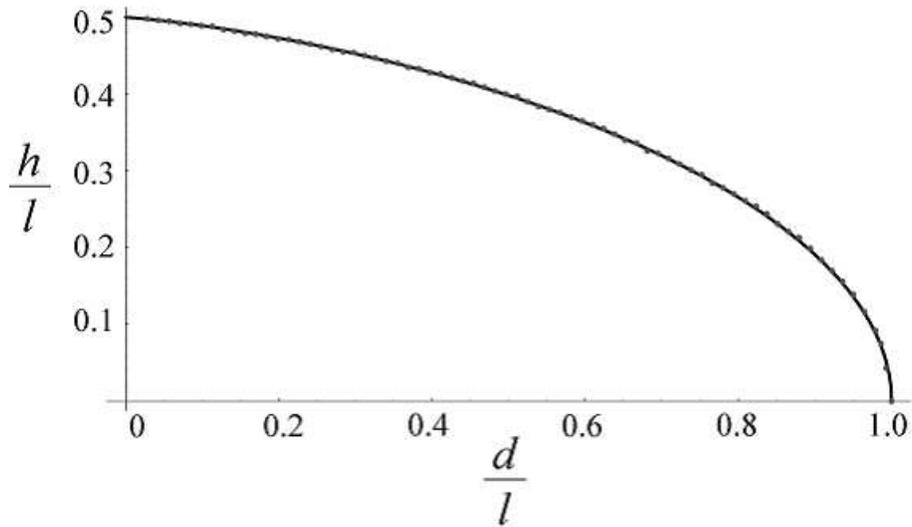}
\end{center}
\caption{For the graph of dimensionless $\dis\frac{h}{l}$ and 
$\dis\frac{d}{l}$,
the dots are the measurement data, and the solid curve is
Eq. (14).}
\end{figure}
\begin{figure}[ht]
\begin{center}
\includegraphics[width=12cm]{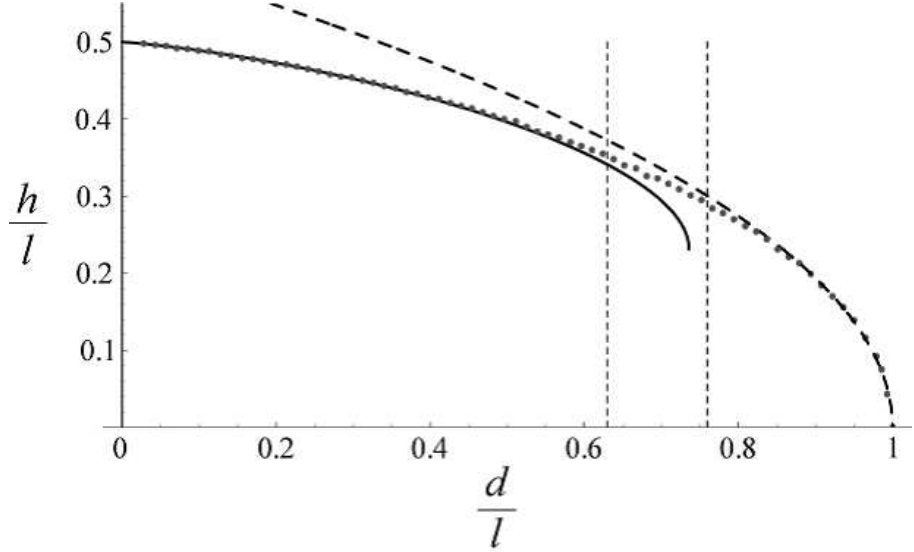}
\end{center}
\caption{For the graph of dimensionless $\dis\frac{h}{l}$ and 
$\dis\frac{d}{l}$,
the broken curve is Eq. (17), and
the solid curve is Eq. (21).
The two vertical broken lines correspond to $\dis\frac{d}{l}=0.63,0.76$.}
\end{figure}
\begin{figure}[ht]
\begin{center}
\includegraphics[width=13cm]{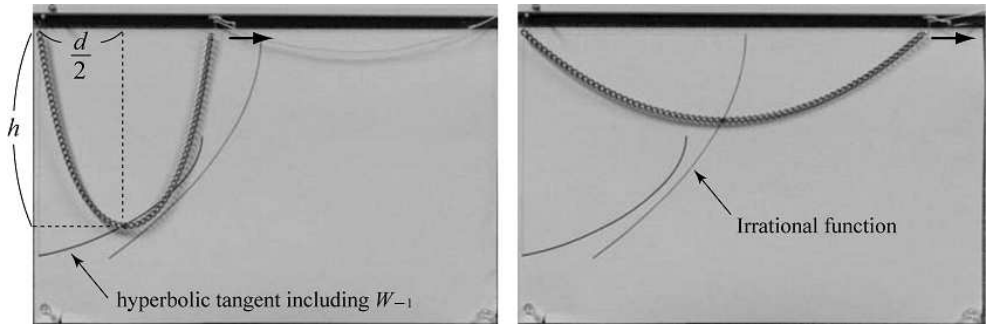}
\end{center}
\caption{The photographs of the demonstration observing the middle point while 
pulling the movable point of the chain from left to right. The length of the chain is $l=40$ cm.
The two solid curves correspond to the plot of $\dis\left(\frac{d}{2},h\right)$ in Eq. (23)}
\end{figure}
\end{document}